\documentstyle[amssymb,aps,preprint]{revtex}
\begin{document}
\draft
\preprint{CALT-68-2061}
\title{Condensate fluctuations of a trapped, ideal Bose gas}
\author{H. David Politzer}
\address{California Institute of Technology\\
Pasadena, California 91125\\
{\tt politzer@theory.caltech.edu}}
\date{June 10, 1996}
\maketitle

\begin{abstract}
For a non-self-interacting Bose gas with a fixed, large number of particles
confined to a trap, as the ground state occupation becomes macroscopic, the
condensate number fluctuations remain microscopic. However, this is the only
significant aspect in which the grand canonical description differs from
canonical or microcanonical in the thermodynamic limit. General arguments
and estimates including some of the vanishingly small quantities are
compared to explicit, fixed-number calculations for 10$^2$ to 10$^6$
particles.
\end{abstract}


%\preprint{CALT-68-2061}

%%INSERT PREPRINT AFTER DRAFT%%

\section{\bf Introduction }

Large fluctuations are a salient feature of the thermal behavior of systems
of bosons. For example, if $n$ is the mean number of non-interacting
particles occupying a particular one-particle state, then the mean-square
occupation fluctuation is $n(n+1)$. This is easily derived in the grand
canonical picture by considering diffusive equilibrium with a particle
reservoir characterized by a chemical potential\cite{kittel}. If, however,
the system has a {\it fixed} total number of particles, $N$, confined in
space by a trapping potential or container, then at low enough temperature $%
T $ or fixed total energy $E$ when a significant fraction of $N$ are in the
ground state, such large fluctuations are impossible. No matter how large $N$%
, this aspect of the grand canonical description cannot be even
approximately true. This paper addresses what {\it does} happen for fixed
large $N$ as $N\rightarrow \infty .$

A decades-old answer to this question is that any interaction between the
particles would eliminate such large fluctuations, even in the presence of a
chemical potential. With a weak inter-particle interaction and a chemical
potential, fluctuations in the occupations of various states are only weakly
correlated. Therefore, the fluctuation in the total number of particles {\it %
not }in the ground state is microscopic. Hence, a macroscopic condensate
fluctuation would mean a macroscopic density fluctuation. Even if the
particles interact weakly, this would mean a macroscopic energy fluctuation.
The consequent macroscopic rise in free energy would suppress the
fluctuation. (See Appendix B for a more formal sketch of this argument.)
Thus, with interactions producing a finite compressibility, the equivalence
of the three standard statistical ensembles is assured in the thermodynamic
limit, and the computationally convenient chemical potential can still be
used for isolated, large systems\cite{huang}. In the context of Bose
liquids, the ideal gas is a theoretical curiosity. Large condensate
fluctuation is only one of several features for which ignoring interactions
gives qualitatively incorrect results\cite{weichman}.

This argument does not address the question of what {\it does} happen to
condensate fluctuations of the ideal Bose gas. Furthermore, this is not a
totally idle or purely theoretical question. In current experimental work on
the trapping and cooling of bosonic atoms, there is typically no diffusive
particle or thermal energy reservoir\cite{wieman,hulet,ketterle}. While the
atoms most certainly interact, $N\neq \infty $. Hence, one can ask about the
system as a whole rather than only describing densities (intensive
quantities), which are really just sub-volumes in diffusive and thermal
equilibrium with their (much larger) surroundings. For sub-volumes of an
infinite system, $\mu $ and $T$ give an appropriate description. However,
for a finite, isolated system taken as a whole, which has a greater impact
on the condensate fluctuations, the particle interactions or the constraint
of fixed total $N$? The answer depends on the density realized in the
particular situation. A practical distinction of a gas from a liquid is that
the density can be easily varied over many orders of magnitude. In the first
successful experiments\cite{wieman}, there are noticeable effects of
interparticle repulsion; and many of the more detailed observations
currently underway require a mean field (albeit weak) description of the
interparticle scattering length to reconcile theory with observations.
Nevertheless, it is possible to imagine approaching Bose condensation with a
box or trap so large and density so low that the effects of a given
inter-atomic interaction, characterized by a fixed scattering length, are
negligible, even for density fluctuations of order the equilibrium density.
(An estimate of the requisite relation of the scattering length, trap
parameters and density is given in Appendix B.) Even though the
Bose-Einstein transition temperature decreases with decreasing density, the
total energy shift due to a weak fixed-strength inter-particle interaction
decreases faster. Also, the actual inter-atomic interactions may not serve
to stabilize anything. Rather, the gaseous state may itself only be
metastable\cite{hulet}. In such situations, the equilibrium statistics of
the ideal gas are certainly a better starting approximation than the
equilibrium statistics of the interacting system.

After a summary of a variety of potentially confusing issues (sec. II), a
thoroughly elementary analysis of the problem (sec. III) suggests that the
condensate fractional fluctuations vanish with increasing $N$, but all other
significant grand canonical predictions have vanishing corrections. This is
also sufficient to establish the equivalence of using either fixed $T$ or
fixed total $E$ to characterize the system for large $N$. The proposed
picture provides an explicit prediction (sec. IV) for the condensate
fluctuation as well as the values of observables, e.g. two-level
correlations, that are identically zero with a chemical potential but are
induced by fixing $N$. (With a natural normalization, such functions are
vanishingly small as $N\rightarrow \infty $.) The results of a numerical
evaluation of the canonical partition function and related functions for $N$
from $10^2$ to $10^6$ (sec. V) confirm these predictions. Some obvious
conclusions are offered (sec. VI), while comments on details of the
numerical work are left to Appendix A. Appendix B outlines the simple
estimate of the condensate fluctuation damping due to repulsive
interactions, which allows a comparison with the effect due to fixing $N$.  

\section{Potential Issues}

It is only the non-interacting particles in the ground state of a trap or
confining potential that do not satisfy the hypotheses of the standard
demonstration\cite{huang} of the equivalence of the grand canonical and
canonical ensembles in the thermodynamic limit. Hence, the questions raised
here only arise if the ground state occupation is macroscopic. At ultra-low $%
T$ when almost all particles are in the ground state, the condensate serves
as a particle reservoir for all the excited states, and so some form of the
grand canonical description for excited states should be valid in that
domain. But what about intermediate $T$'s? Is the inequivalence of chemical
potential and fixed $N$ limited to the size of the ground state
fluctuations? If the condensate manifested the boson propensity for large
fluctuations and there were {\it any} macroscopic fluctuation in the
condensate number, it would have to be accompanied by correlations between
the various occupation numbers. (Such correlations are identically zero for
the grand canonical ideal gas.) There need not be any macroscopic
fluctuation in the average density because the total number is fixed. Yet,
larger than anticipated exited state fluctuations and correlations might
lead to larger fluctuations in the total $E$ at fixed $T$. And were this the
case, the equivalence of fixing $E$ and fixing $T$ might be lost in the
thermodynamic limit.

Chemical potential is not just a calculational convenience. There is really
no practical alternative for analytic calculations because not much is known
directly about the large but fixed $N$ asymptotics of the canonical or
microcanonical partition functions, even for systems as simple as the ideal
Bose gas. If this analytic tool were lost, theory would be reduced almost
entirely to numerical techniques.

\section{Fixed-$N$ Statistics}

The resolution of these conundrums lies in the observation that the grand
canonical excited state occupations in the thermodynamic limit are
independent of not only the condensate fluctuations but the condensate
occupation itself. Hence, if the behavior of the excited state occupancies
can be reliably estimated using the concept of a chemical potential, one can
deduce the behavior of the condensate from the constraint of fixed $N$. This
argument is really just a minor extension of the traditional one used to
compute the condensate fraction\cite{london,kittel}. In particular, it goes
as follows.

Let $i$ label the one-particle (or trap) states and $\varepsilon _i$ be
their energies. Take $i=0$ to be the lowest energy level, and take $%
\varepsilon _i=0$. In the presence of a chemical potential $\mu $, the mean
occupation numbers $N_i$ for non-interacting bosons are 
\begin{equation}
N_i=\frac 1{e^{(\varepsilon _i-\mu )/T}-1}\text{ .}  \label{N_i occupation}
\end{equation}
With the chosen zero of energy, 
\begin{eqnarray}
e^{-\mu /T} &=&1+\frac 1{N_0}  \nonumber \\
&\equiv &\lambda ^{-1}\text{ }  \label{fugacity}
\end{eqnarray}
(defining the fugacity $\lambda $, to be used later). Once $N_0\gg 1$ (which
may still be for $N_0\ll N$), the explicit fixed-$T$ $N_0$ dependence of $%
N_{i>0}$ is ${\cal O}(1/N_0)$. The expression for the expected total number
of particles with $i>0$, $N_e$, and how it depends on $\mu $ is determined
by the density of states. For an isotropic harmonic oscillator potential in
three dimensions with level spacing $\epsilon $, 
\begin{equation}
N_e=\zeta (3)\text{ }(T/\epsilon )^3  \label{Ne}
\end{equation}
as long as $N_e<N$ and $T/\epsilon \gg 1\cite{deGroot}.$ Under the latter
condition, the asymptotic behavior of the sum over states is given by an
integral. ($\zeta (3)\approx 1.202$ is the Riemann Zeta function.) Under
these circumstances, the fixed-$T$ corrections to eq. (\ref{Ne}) are ${\cal O%
}(1/N_0)$. The root-mean-square fluctuation of any occupation number is
precisely 
\begin{equation}
\Delta N_i=\sqrt{N_i(N_i+1)}\text{ .}  \label{delta Ni}
\end{equation}
For the isotropic oscillator, this implies 
\begin{equation}
\Delta N_e=\sqrt{\frac{\pi ^2}6(T/\epsilon )^3}\text{ ;}  \label{delta Ne}
\end{equation}
so $\Delta N_e/N_e\sim {\cal O}(1/\sqrt{N_e})$. The corrections to eq. (\ref
{delta Ne}) for $\mu $ not exactly zero are again ${\cal O}(1/N_0)$.

The success of using a $\mu $ to characterize a system with a large but
fixed total number of particles $N$ relies on the fact that each individual
energy level is a system in diffusive equilibrium with the much larger
remainder of the total system. This remainder acts as the single level's
particle reservoir. Once $N_0$ is not much less than $N$, the utility of $%
\mu $ is no longer clear. Certainly there exists no yet-much-larger particle
reservoir for the ground state.

Referring back to eq. (\ref{N_i occupation}), once $N_0$ is large, the only
role of the particular value of $\mu $ is to determine $N_0$. The $N_{i>0}$
are insensitive to $\mu $ or $N_0$. So, if we consider each individual
excited level with $i>0$ as a system in contact with the reservoir of all
the other levels, we need not know exactly what the chemical potential
actually is, only that it is nearly zero. In fact, there need not be any
precise meaning to $\mu $, only that it is nearly zero. It may be impossible
to disentangle the effects of ``$\mu \neq 0"$ from other $1/N$ consequences
of fixing the total $N$. From this perspective, $N_0$ is determined not by a 
$\mu $ but by $N$ and $N_e$: 
\begin{equation}
N_0=N-N_e\text{ .}  \label{N cons.}
\end{equation}
However, this is precisely the same value of $N_0$ that is deduced from eq. (%
\ref{N_i occupation}) when $N$ is interpreted as an expectation in the
presence of an external $\mu $.

At the level of occupation expectations, the assignments given by eq. (\ref
{N_i occupation}) for $i>0$ minimize the total free energy (energy minus $%
T\times $entropy) irrespective of the actual value of $N_0$ or $N$ as long
as $N_e$ is fixed. This is because adding or removing particles from the $%
i=0 $ condensate changes neither the energy nor the entropy of the entire
system. Hence, for large $N_0$, the occupation numbers for $i>0$ are
unchanged from their grand canonical values if, instead of being determined
by a diffusive equilibrium, $N$ is fixed at some value and $N_0$ is large.
Once there is a condensate, the only thing that can change as particles are
added at fixed $T$ is $N_0$.

The total expected energy $\langle E\rangle $ at fixed $T$ depends only on
the $i>0$ occupations. Thus canonical and grand canonical evaluations of the
total energy must agree as $N\rightarrow \infty $. For the isotropic
harmonic trap 
\begin{eqnarray}
\langle E\rangle &=&\frac{\pi ^4}{30}T^4\epsilon _{}^{-3} \\
&=&\frac{\pi ^4}{30\zeta (3)}TN_e\text{ .}  \nonumber
\end{eqnarray}
Since it is a canonical ensemble identity that the root-mean-square total
energy fluctuation satisfies 
\begin{equation}
\Delta E=\sqrt{T^2\frac{\partial \langle E\rangle }{\partial T}}\text{ ,}
\end{equation}
the equivalence of the canonical and microcanonical ensembles is assured as
long as $N_e\rightarrow \infty $ because $\Delta E/E\sim {\cal O}(1/\sqrt{N_e%
})$. (This is true for any trapping potential, not just the explicit example
given.)

\section{Fluctuation Estimates}

From the discussion above, it is expected that all occupations approach
their grand canonical values as $N\rightarrow \infty $, even if either or
both $N$ and $E$ are fixed. One can go further and estimate the leading
behavior of various quantities that vanish in this limit. As examples I
consider the condensate fluctuations and the occupation correlations between
levels.

As long as $N_0\ll N$, the root-mean-square fluctuation in the condensate
number, $\Delta N_0$, satisfies eq. (\ref{delta Ni}). Once $N_0\sim {\cal O}%
(N)$, eq. (\ref{N cons.}) implies 
\begin{equation}
\Delta N_0=\Delta N_e\text{ .}  \label{delta N0}
\end{equation}
The cross over between these two behaviors is an example of the phenomena
that make a direct analysis of the fixed-$N$ partition function difficult.
It is appropriate to introduce the ``critical" temperature $T_c$, given by
the point at which $N_e$ reaches $N$ or, rather, at which $N_0$ goes from
macroscopic to microscopic. For the isotropic harmonic potential, eq.(\ref
{Ne}) implies 
\begin{equation}
T_c=N^{1/3}\zeta (3)^{-1/3}\epsilon \text{ .}  \label{Tc}
\end{equation}
As $N$ increases, $T_c$ remains fixed in absolute, physical units only if
the trap size is increased, e.g. $\epsilon $ decreased. The transition
occurs when the central density in the trap reaches the infinite volume
critical value\cite{bagnato}. In terms of the natural temperature variable
for the study of Bose-Einstein condensation, $T/T_c$, the transition between
eq. (\ref{delta Ni}) and eq. (\ref{delta N0}) takes place in a vanishingly
small interval as $N\rightarrow \infty $.

In the thermodynamic limit with $N$, $N_0$, and $N_e$ all very large, eqs. (%
\ref{Ne},\ref{delta Ne},\ref{N cons.},\ref{delta N0},\ref{Tc}) can be
combined to give a simple estimate of the leading behavior: 
\begin{equation}
\frac{\Delta N_0}{N_0}=\frac 1{\sqrt{N}}\frac{(T/T_c)^{3/2}}{1-(T/T_c)^3}%
\left( \frac{\pi ^2}{6\zeta (3)}\right) ^{1/2}\text{ .}
\label{limit delta N0}
\end{equation}

A set of quantities that are of interest in the calculation of the angular
dependence of light scattering off cold, trapped atoms \cite{politzer} are
the two-level occupation expectations, $\langle n_in_j\rangle $. (I use the
notation ``$n_i"$ for the actual $i$th level occupation number in a
particular configuration of the thermal ensemble.) In the grand canonical
analysis of an ideal Bose gas, these are given precisely by $N_iN_j$. In
particular, there is no correlation between the fluctuations in one level
and another. However, with $N$ fixed, this cannot be exactly true. A
refinement of the argument of the previous section allows one to estimate
the leading behavior of these correlations. As an example consider the two
states with the largest fluctuations, $i=0$ and $j=1$, because their fixed-$%
N $ induced correlation must, therefore, be the largest:

At fixed $N$, if $n_0$ fluctuates down, say, then $n_e$ must fluctuate up by
an equal amount. The impact on the $n_{i>0}$ can be estimated by computing
the particular expected $N_i$ given that $N_e$ is larger than its original
equilibrium value by the negative of the $i=0$ fluctuation. This implies
(writing $\Delta n_i$ for $n_i-\langle n_i\rangle $) 
\begin{equation}
\langle \Delta n_0\Delta n_1\rangle =\langle \Delta n_0^2\rangle \frac{%
\delta \langle n_1\rangle }{\delta \langle n_0\rangle }=\langle \Delta
n_0^2\rangle \left( -\frac{dN_1/d\lambda }{dN_e/d\lambda }\right) _{\lambda
=1}\text{ .}  \label{correl'}
\end{equation}
The fugacity, $\lambda $, is defined by eq.(\ref{fugacity}). For the
isotropic, harmonic trap in the thermodynamic limit, this can be evaluated
to give (with the natural normalization factor $N_0N_1$) 
\begin{equation}
\frac{\langle \Delta n_0\Delta n_1\rangle }{N_0N_1}=-N^{-2/3}\frac{T/T_c}{%
1-(T/T_c)^3}\zeta (3)^{-1/3}\text{ . }  \label{correl}
\end{equation}

\section{Numerical Evaluation of the Canonical Ensemble}

The canonical partition function, $Z(N,T)$, of a trapped, ideal Bose gas can
be represented as 
\begin{equation}
Z(N,T)=\frac 1{2\pi }\int_{-\pi }^\pi dz\text{ }e^{iNz}\prod_{m=0}^\infty
\sum_{n_m=0}^\infty e^{-n_m\varepsilon _m/T-in_mz}  \label{Z}
\end{equation}
where $n_m$ is the number of particles in the state labeled by $m$ with
energy $\varepsilon _m$. The integral over $z$ implements the constraint $%
N=\sum_mn_m$. For the isotropic harmonic potential in three dimensions, it
is convenient to let $m$ label the energy {\it levels} $\varepsilon
_m=m\epsilon $, with the associated degeneracy of $\frac 12(m+1)(m+2)$ for $%
m=0,1,2,$.... The infinite sums over occupations can be done explicitly.
Occupation expectations and correlations can be represented similarly by
simple modifications of the integrand, i.e. extra weight factors of $n_i$ or 
$n_in_j$. If one truncates the infinite product over energy levels $m$ at
some finite $M_{\max }$, this yields a form that can be evaluated
numerically. One can study the convergence in $M$ to test whether the
asymptotic values of thermal expectations have plausibly been reached.
[Useful numerical strategies and some details of the evaluations are
provided in Appendix A.]

Fig. (\ref{fig1}) shows the results of calculations of $N_0$. In particular,
the solid lines are the numerically computed values of $N_0/N$ for $N=10^2$, 
$10^3$, $10^4$, and $10^6$, plotted versus $T/T_c$, where $T_c$ is given by
eq.\ (\ref{Tc}) appropriate to each $N$. The dotted lines are grand
canonical predictions for $N=10^2$ (small dots) and the $N\rightarrow \infty 
$ limit, $1-(T/T_c)^3$ (large dots). Note that the grand canonical
predictions were computed as {\it sums} over states using eqs. (\ref
{N_i
occupation},\ref{fugacity}) and involve no approximations regarding $N$%
. The comparison of the two statistical ensembles is made by identifying the
value of the grand canonical $\langle N\rangle $ with the precise canonical $%
N$. The canonical numerical calculations clearly approach the $N\rightarrow
\infty $ grand canonical form as a limiting value with increasing $N$. For
intermediate values of $T/T_c$, e.g. $0.6$, the {\it fractional }discrepancy
between the canonical $N$ and $N\rightarrow \infty $ , i.e.
difference-divided-by-value, appears to be decreasing roughly like $%
N^{-0.33} $.

The differences between canonical and grand canonical values for $N_0$ are
displayed in another way in fig. (\ref{fig2}). The fractional discrepancy
between the two evaluations are plotted for $N=10^2$, $10^3$, and $10^4$
versus $T/T_c$. Here, ``fractional discrepancy'' means $(N_0^{grand\text{ }%
canonical}-N_0^{canonical})/N_0^{grand\text{ }canonical}$. At very small $%
T/T_c$, all evaluations give $N_0/N$ very near to $1$. So the ratio plotted
in fig. (\ref{fig2}) plummets, but it is not an effective way to
characterize the difference between fixed $N$ and fixed $\mu $. (For that
region, a more informative variable would be $N_1$.) For intermediate values
of $T/T_c$, the curves of fig. (\ref{fig2}) decrease roughly like $N^{-1.15}$%
. So, not only does the canonical $N_0$ approach $N(1-(T/T_c)^3)$ as $%
N\rightarrow \infty $, it does so approximately as predicted by the simple
grand canonical calculation. It is the next correction, the difference
between the two ensembles' predictions at a given $N$ (as illustrated in
fig. (\ref{fig2})) that reflects the residual difference in physics between
the ensembles. This difference is particularly pronounced as $N_0$ makes the
transition from micro- to macroscopic just below $T_c$. There, the grand
canonical -- canonical discrepancy decreases only very slowly with $N$. The
sign and shape of the difference is such that the canonical $N_0$ does not
rise quite as sharply as the grand canonical, but the width of the relevant
region of $T/T_c$ vanishes with increasing $N$. Above $T_c$, the distinction
between fixing $N$ and fixing $\mu $ has rapidly vanishing consequences.

The dashed lines in fig. (\ref{fig1}) are the results of a numerical
evaluation of the canonical $\Delta N_0/[N_0(N_0+1)]^{1/2}$ versus the same $%
T/T_c$'s for $N=10^2$, $10^3$, and $10^4$. For $T\gtrsim T_c$, this ratio
approaches $1$, in agreement with the grand canonical eq. (\ref{delta Ni}).
However, for $T<T_c$, it goes to zero, more dramatically with increasing $N$%
. This same $\Delta N_0$ data is plotted again on a log scale as the solid
lines in fig. (\ref{fig3}). The dotted lines are plots of eq. (\ref
{limit
delta N0}) for the same $N$'s. As long as neither $N_0$ nor $N_e$
are too small, eq. (\ref{limit delta N0}) clearly captures the $N$ and $T$
dependence of $\Delta N_0$, and the agreement improves with increasing $N$.
In particular, the fractional discrepancy between the canonical and eq. (\ref
{limit delta N0}) values appears to go roughly like $N^{-0.25}$.

The canonical, normalized, fluctuation correlation, $-\langle \Delta
n_0\Delta n_1\rangle /N_0N_1$, is plotted (solid lines) on a log scale
versus $T/T_c$ for $N=10^2$, $10^3$, and $10^4$ in fig. (\ref{fig4}). The
overall minus sign is because the correlation is, indeed, negative. The
dotted lines are eq. (\ref{correl}) for the same three $N$'s, and again the
agreement improves with $N$; this time the fractional discrepancy appears to
go roughly like $N^{-0.33}$.

The discrepancies between the numerical evaluations and the simple formulae
are largest for $T$'s such that either $N_0$ or $N_e$ are not very large.
These are vanishingly small intervals of $T/T_c$ for $N\rightarrow \infty $.

The expected $i=1$ occupation, $N_1$, was evaluated to prepare fig.\ (\ref
{fig4}). The agreement with eq. (\ref{N_i occupation}) with $\mu =0$ was
such that the leading fractional discrepancy was accounted for by just the
leading $1/N_0$ correction already included in eq. (\ref{N_i occupation}),
i.e. $T/N_0$.

The particular computer code used for the results presented was checked
against hand calculations for small $N$. For large $N$, a criterion for
validity was stability under changes in the several parameters that should
not effect the final answers. Eventually, at high enough $N$ (different
values for different observables) the ranges of stability in these
parameters shrunk to zero. Typically, the practical limitation was the
digits of precision available for intermediate results. The code was written
to evaluate $N_0$ below $T_c$, and specifically for $N_0$ plausible results
were obtained for much higher $N$ than presented. No effort was made to
modify the numerical strategy to facilitate calculation of the other
quantities discussed; presumably those calculations could be extended to
higher $N$ with algorithmic improvements that avoided the simultaneous
evaluation of numbers of vastly different magnitudes.

\section{Discussion and Conclusions}

The general arguments presented here, while heuristic, have an internal
consistency. For example, to compute $\Delta N_e$, which is used implicitly
in eqs. (\ref{correl'},\ref{correl}), one assumes that the correlations
between level occupations are negligible. One then deduces non-zero
correlations that are induced by particle conservation. However, the induced
correlations are, indeed, small enough to be ignored in the calculation of
the leading behavior of $\Delta N_e$ and of the correlations themselves.

This is nowhere near to a ``theory'' of the large $N$ asymptotics of the
canonical ideal Bose gas. The leading behavior of some interesting
observables were estimated and confirmed numerically. But in these cases,
the leading behavior either was simply given by or could be deduced from the
grand canonical ensemble. The next level of approximation, e.g. to account
for fig. (\ref{fig2}), would require a detailed analysis of the canonical or
microcanonical partition function and may be very difficult to determine
analytically.

Starting with the grand canonical description with $\mu$ and $T$ as
independent variables, one finds large fluctuations in $N$ below $T_c$.
Hence, fixing $N$ may have been expected to be of some consequence. However,
the grand canonical total energy fluctuations are always small and vanish
relative to the mean total energy in the thermodynamic limit. Nothing
special happens in $E$ at $T_c$. So fixing $E$ should have no dramatic
consequences. Overall, the switch from $T$ to $E$ should be of even less
consequence than the switch from $\mu$ to $N$. A direct numerical evaluation
of the microcanonical partition function would be considerably more
difficult.

However, from a practical standpoint, the modest results here are useful.
The largest consequence of going from a chemical potential to fixed $N$ for
an ideal Bose gas is that the ground state number fluctuations are always
microscopic; the leading behavior of all expected level occupations are
unchanged. This is sufficient to further imply that fixing the total $E$ is
no different from the analytically simpler fixing of $T$ in the
thermodynamic limit. The leading behaviors two-level expectations, $\langle
n_in_j\rangle $ for $i\neq j$, are unchanged because the induced
correlations vanish as $N\rightarrow \infty $. For large, fixed $N$, the
corrections to these behaviors are unlikely to be of any practical
importance. As discussed in Appendix B, for a gas with replusive
interactions, the consequence of fixing $N$ dominates over the interaction
effects in damping the ground state number fluctuations only if the pairwise
energy in the ground state is less than ${\cal O}(N^{-2/3}\epsilon ).$

\acknowledgements Werner Krauth of E.N.S., Paris, pointed out that an
earlier effort along these lines was in error and suggested the numerical
strategy followed here. Anton Kapustin patiently offered suggestions and
criticism. This work was supported in part by the U.S. Dept. of Energy under
Grant No. DE-FG03-92-ER40701.

\section*{appendix a: numerical strategies}

For the isotropic harmonic potential in three dimensions and a maximum
energy level $M_{\max }$, eq. (\ref{Z}) takes the explicit form 
\begin{equation}
Z(N,T)=\frac 1{2\pi }\int_{-\pi }^\pi dz\text{ }e^{iNz}\prod_{m=0}^{M_{\max
}}\left( \frac 1{1-e^{-m\epsilon /T-iz}}\right) ^{\frac 12(m+1)(m+2)}\text{.}
\label{Za}
\end{equation}
A rather primitive C program on a Sun SPARC10 for integrating eq. (\ref{Za})
and related functions was sufficient to generate the numerical results
presented in the figures, with the size of $N$ limited by the use of
double-precision arithmetic. A few general observations may prove to be of
some value in any future effort to perform comparable calculations.

Instead of simply truncating the product over energy levels $m$ at some
large value $M_{\max }$, one can use Maxwell--Boltzmann statistics for all
levels $m>M_{\max }$ and derive an approximate closed form for the
contribution to the integrand of all levels above $M_{\max }$. This vastly
improves the rate of convergence in $M_{\max }$ because for modest $m$'s
(e.g. $6\times T/\epsilon $) there are still quite a few particles at that $%
m $ or higher, but the occupations of individual states are rarely greater
than $1$.

By far the most rapid variation of the integrand for large $N$ comes from
the factor $e^{iNz}$. The integration algorithm should reflect this
knowledge. For example, one can divide $z$ into intervals of $\pi /4N$ and
integrate each interval accordingly. (For the largest of $N$'s it proved
sufficient to take a single point in each such interval.)

An overall factor in $Z$ has no effect on physical observables. This can be
used to considerable advantage. Here are a couple of examples: One can
evaluate the products of very large numbers logarithmically, i.e. sum the
phase and log(modulus) of the various complex factors. An overall shift
before exponentiation and addition (integration) keeps numbers from getting
too big. Also, observables are independent of shifts of the whole energy
spectrum by the ground state energy $\varepsilon _0$. It is convenient to
take this non-zero to check the numerical independence. Taking $\varepsilon
_0\neq 0$ can also dramatically alter the character of the integrand of eq. (%
\ref{Za}) --- note the (analytically integrable) singularity at $z=0$ for $%
\varepsilon _0=0$.

It is, of course, sufficient to integrate only $0\leq z\leq \pi $. With
suitable choice of $\varepsilon _0$, starting at $z=0$ one can integrate
outward, test the convergence, and exit the integration long before reaching 
$z=\pi $.

\section*{appendix b: interaction damping of grand canonical occupation
fluctuations}

The effect of a weak repulsive interaction on condensate fluctuations can be
estimated as follows. Let $n_0$ represent the number of particles in the
ground state. The leading effect of a weak, pairwise repulsion at low $T$,
when most of the particles are in the ground state, is to raise the energy
of those particles from $0$ (a convenient $n$-independent normalization of
the non-interacting ground state energy) to $\lambda n_0^2$, where $\lambda $
is the positive two-particle interaction contribution to the ground state
energy. In natural oscillator units ($\hbar =m=\omega _0=1$), $\lambda $ is
related to the conventionally defined scattering length $a$ by $\lambda =a/%
\sqrt{2\pi }$, at least if the interaction effects are weak enough to be
treated in mean field theory. Focus on the terms in the grand canonical
partition function that refer only to the ground state:

\begin{equation}
{\cal Q}_0(\mu ,T)=\sum\limits_{n_0=0}^\infty e^{\mu n_0/T-\lambda n_0^2/T}%
\text{ .}
\end{equation}

\noindent Unlike the $\lambda =0$ case, one can now get large $N_0$ ($%
=\langle n_0\rangle $) with $\mu >0$. Then, the sum can be considered as an
integral over $n_0$, whose integrand is a Gaussian peaked at $n_0=N_0=\mu
/2\lambda $ with width $\Delta N_0=\sqrt{T/\lambda }$. Hence, $\Delta
N_0/N_0=\sqrt{T/(\lambda N_0^2)}$, in contrast to the $\lambda =0$
situation, in which $\Delta N_0/N_0$ is ${\cal O}(1).$ So grand canonical
condensate number fluctuations are small if the interaction contribution to
the ground state energy is large compared to the temperature.

If sufficiently strong, interatomic repulsion will certainly be effective at
damping condensate fluctuations at fixed $N$, giving  $\Delta N_0\sim \sqrt{%
T/\lambda }$. This effect will dominate (i.e. enforce {\it smaller}
fluctuations) over the non-interacting $\Delta N_0\sim \sqrt{(T/\epsilon )^3}
$ estimated in sections III and IV when $\lambda /\epsilon \gtrsim
(T/\epsilon )^{-2}$. ($\epsilon $ is the trap level spacing, and $%
T_c/\epsilon \sim N^{1/3}$.)

\begin{figure}[tbp]
\caption{Canonical $N_0/N$ for $N=10^2,10^3,10^4,$ and $10^6$ (solid lines),
grand canonical $N_0/N$ for $N=10^2$ (small dots), the grand canonical $%
N\rightarrow \infty $ limit (large dots), and the normalized canonical
condensate RMS fluctuations (dashed lines) for $N=10^2,10^3,$ and $10^4$
vesus $T/T_c$.}
\label{fig1}
\end{figure}

\begin{figure}[tbp]
\caption{Comparison of the canonical and grand canonical values for $N_0$ as
fractional discrepancies on a log scale for $N=10^2,10^3,$ and $10^4$ versus 
$T/T_c$.}
\label{fig2}
\end{figure}

\begin{figure}[tbp]
\caption{A log plot of the canonical condensate RMS fluctuations (solid
lines) and the simple eq. (\ref{limit delta N0}) estimates (dotted lines)
for $N=10^2,10^3,$ and $10^4$.}
\label{fig3}
\end{figure}

\begin{figure}[tbp]
\caption{A log plot of -1 times the normalized, canonical 0--1 level
correlations (solid lines) and the simple eq. (\ref{correl}) estimates
(dotted lines) for $N=10^2,10^3,$ and $10^4$.}
\label{fig4}
\end{figure}

\end{document}